\documentclass[twocolumn,conference,letterpaper]{IEEEtran}
\usepackage[left=0.75in,right=0.75in,top=1in,bottom=0.75in]{geometry}

\usepackage{cite}
\usepackage{amssymb,amsmath,latexsym,amsfonts,amsthm,mathtools}

\usepackage{graphicx}
\usepackage{float,subcaption}
\usepackage{textcomp}
\usepackage{xcolor}
\definecolor{darkpastelpurple}{rgb}{0.59, 0.44, 0.84}
\usepackage{hyperref}
\hypersetup{colorlinks, breaklinks, citecolor=blue, linkcolor=blue, urlcolor=blue}
\usepackage[nameinlink]{cleveref}
\Crefformat{figure}{#2Fig.~#1#3}
\Crefmultiformat{figure}{Figs.~#2#1#3}{ and~#2#1#3}{, #2#1#3}{ and~#2#1#3}

\usepackage{nicematrix}

\usepackage{tikz}
\usetikzlibrary{automata, shapes, arrows, calc, arrows.meta, fit, positioning}

\usepackage{tkz-graph}
\GraphInit[vstyle = Normal]
\SetUpVertex[FillColor=blue!30]  
\tikzset{
	LabelStyle/.style = { rectangle, rounded corners, draw,
		minimum width = 2em, fill = green!30,
		text = black, font = \bfseries },
	VertexStyle/.append style = { inner sep=5pt, text=black	},
	EdgeStyle/.append style = {->,>=stealth',shorten >=0pt, bend left} }

\theoremstyle{plain}

\newtheorem{lemma}{Lemma}

\newtheorem*{problem*}{Problem}
\theoremstyle{remark}
\newtheorem{remark}{Remark}

\newtheorem{definition}{Definition}
\theoremstyle{definition}

\usepackage{mathrsfs}

\usepackage{newtxmath}
\usepackage{booktabs}
\usepackage{duckuments}
\usepackage{bm}

\IEEEoverridecommandlockouts

\begin{document}
	\title{On Robustness of Consensus over Pseudo-Undirected Path Graphs}
	\author{Abhinav Sinha,~\IEEEmembership{Senior Member,~IEEE}, Dwaipayan Mukherjee,~\IEEEmembership{Member,~IEEE},\\ and Shashi Ranjan Kumar,~\IEEEmembership{Senior Member,~IEEE}
		\thanks{A. Sinha is with the GALACxIS Lab, Department of Aerospace Engineering and Engineering Mechanics, University of Cincinnati, OH 45221, USA. (email: abhinav.sinha@uc.edu).  D. Mukherjee is with the Department of Electrical Engineering, Indian Institute of Technology Bombay, Powai- 400076, Mumbai, India, e-mail: dm@ee.iitb.ac.in. S.R. Kumar is with the Intelligent Systems \& Control Lab, Department of Aerospace Engineering, Indian Institute of Technology Bombay, Powai- 400076, Mumbai, India, e-mail: srk@aero.iitb.ac.in.}
	}

	\maketitle
	\thispagestyle{empty}
	
	\begin{abstract}        
        Consensus over networked agents is typically studied using undirected or directed communication graphs. Undirected graphs enforce symmetry in information exchange, leading to convergence to the average of initial states, while directed graphs permit asymmetry but make consensus dependent on root nodes and their influence. Both paradigms impose inherent restrictions on achievable consensus values and network robustness. This paper introduces a theoretical framework for achieving consensus over a class of network topologies, termed pseudo-undirected graphs, which retains bidirectional connectivity between node pairs but allows the corresponding edge weights to differ, including the possibility of negative values under bounded conditions. The resulting Laplacian is generally non-symmetric, yet it guarantees consensus under connectivity assumptions, to expand the solution space, which enables the system to achieve a stable consensus value that can lie outside the convex hull of the initial state set. We derive admissibility bounds for negative weights for a pseudo-undirected path graph, and show an application in the simultaneous interception of a moving target. 
	\end{abstract}
	
	
	\section{Introduction}\label{sec:introduction}
 Graph-theoretic models play a central role in the analysis of distributed dynamical systems, with consensus problems serving as a foundational setting \cite{mesbahi2010graph}. In such problems, agents exchange information according to a communication graph and iteratively update their states with the goal of reaching agreement. The properties of the underlying graph, whether directed or undirected, determine both the convergence behavior and the eventual consensus value. Two major paradigms have been extensively studied. On one hand, undirected graphs enforce symmetry in information exchange, which guarantees convergence to the average of initial states. This symmetry yields elegant algebraic properties-- the Laplacian matrix is symmetric, positive semidefinite, and its eigenvalues form a well-understood spectrum tied to connectivity. However, the same symmetry imposes rigid limitations. In particular, the consensus value is restricted to the convex hull of initial states, and individual nodes cannot exert asymmetric influence. In contrast, directed graphs abandon the strict symmetry of undirected networks by permitting information flow to be inherently asymmetric. Consensus is then governed by the left nullspace of the generally non-symmetric Laplacian, leading to weighted averages where influence can be unevenly distributed. Yet this flexibility comes at a structural cost. Consensus requires the existence of a globally reachable node (or set of roots), making the consensus process vulnerable to over-reliance on specific agents and reducing robustness to variations in edge weights.

 This motivates us to develop an intermediate construct of \emph{pseudo-undirected graphs} \cite{11018241}. In such graphs, every edge appears in a bidirectional pair, preserving the intuitive notion of mutual communication, but the weights in opposite directions may differ. Unlike undirected graphs, the Laplacian is no longer symmetric. Unlike arbitrary directed graphs, the underlying bidirectional structure prevents isolation of influence. Pseudo-undirected graphs thus combine the connectivity properties of undirected graphs with the flexibility of directed and asymmetric weighting. Within this framework, the path graph emerges as a special case because classical path graphs are among the most studied structures in algebraic graph theory due to their simplicity and analytical tractability. Their simple topology enables explicit characterization of Laplacian eigenvalues and eigenvectors, facilitating rigorous analysis of consensus dynamics, diffusion processes, and stability in networked systems. Path graphs also provide clear intuition for how local interactions propagate through a network and shape global behavior \cite{chung}. 
 
 By generalizing these structures to pseudo-undirected path graphs, we introduce asymmetry in edge weights and study the extent to which an edge weight can be perturbed without losing consensus. This generalization allows exploration of new phenomena, including weighted influence distributions, consensus points outside the convex hull of initial states, and altered spectral properties. The concept of assigning heterogeneous weights \cite{10472130,9919791}, including the use of negative edge weights, has demonstrated promising applications (e.g., \cite{8638852,XIAO200465,XIA20112395}). However, the presence of negative weights introduces potential challenges, such as impacts on consensus robustness, that must be carefully analyzed and addressed.  
 
 In this paper, we focus our attention on pseudo-undirected path graphs and derive conditions for achieving consensus over such graphs with heterogeneous (possibly negative) edge weights. We first present a systematic method to compute the left eigenvector corresponding to the null space of the graph Laplacian and then obtain the consensus value. Then, we establish robustness bounds for negative weights to ensure consensus is maintained under perturbations on a single edge weight. Finally, we demonstrate how weighted interactions and deliberate perturbations can be used to tailor the consensus variable, with applications to cooperative guidance and simultaneous interception.

\section{Preliminaries and Problem Formulation}
We denote the set of real numbers by $\mathbb{R}$ and the positive integers by $\mathbb{Z}_+$. Vectors are represented by bold lowercase letters. The column vectors $\vmathbb{1}_p \in \mathbb{R}^p$ and $\vmathbb{0}_p \in \mathbb{R}^p$ contain ones and zeros, respectively. For a matrix $\mathbf{M} = [m_{ij}] \in \mathbb{R}^{p \times p}$, $\lambda_k(\mathbf{M})$ denotes its $k$-th eigenvalue ($k = 1,\ldots,p$), $\mathrm{diag}(\cdot)$ denotes a diagonal matrix with the specified entries, and $\mathbf{I}_p$ is the $p$-dimensional identity. Absolute value and 2-norm are denoted $|\cdot|$ and $\|\cdot\|$, respectively. The convex hull of a set $\mathfrak{C} \subset \mathbb{R}$ is $\mathcal{C}o\{\mathfrak{C}\}$. For a matrix $\mathbf{M}$, $\mathfrak{N}(\mathbf{M})$ and $\mathcal{R}(\mathbf{M})$ denote its null and range spaces, $\rho(\mathbf{M})$ its rank, $\mathfrak{s}\{\cdot\}$ the span of a set of vectors, and $\mathfrak{b}\{\cdot\}$ the corresponding basis.

An undirected simple graph \cite{mesbahi2010graph} is an ordered pair $\mathcal{G} = (\mathcal{V},\mathcal{E})$, where $\mathcal{V} = \{v_1, \dots, v_n\}$ is a finite set of nodes and $\mathcal{E} \subset \{{v_i,v_j} \mid v_i,v_j \in \mathcal{V}, i \neq j\}$ is the set of edges. The cardinalities of $\mathcal{V}$ and $\mathcal{E}$ are $n$ and $m$, respectively, though in pseudo-undirected graphs the number of edges is always even and considered as $2m$. Nodes $v_i$ and $v_j$ are adjacent if $\{v_i,v_j\} \in \mathcal{E}$. The degree of node $v_i$, $d(v_i)$, counts edges incident to it. The degree matrix $\mathcal{D}(\mathcal{G})$ is diagonal with $d(v_i)$ on the $i$-th entry, and the adjacency matrix $\mathcal{A}(\mathcal{G})$ has $[a_{ij}]$ if $\{v_i,v_j\}\in \mathcal{E}$ and $0$ otherwise. The graph Laplacian is $\mathcal{L}(\mathcal{G}) = \mathcal{D}(\mathcal{G}) - \mathcal{A}(\mathcal{G})$. An edge $\{v_i,v_j\}$ is incident on a node $v$ if $v = v_i$ or $v=v_j$. Assigning arbitrary orientations to edges allows defining the incidence matrix $E(\mathcal{G}) \in \mathbb{R}^{n\times m}$, with entries $\pm 1$ if edge $e_j$ is incident on node $v_i$ ($+1$ if oriented away, $-1$ otherwise) and $0$ otherwise \cite{mesbahi2010graph}. A path is a sequence of $k$ consecutively adjacent nodes with length $k-1$. The distance between nodes $v_i$ and $v_j$ is the length of the shortest path connecting them, and the diameter $d$ of the graph is the largest such distance across all node pairs. For a directed graph, the edge set is $\mathcal{E} \subset \mathcal{V} \times \mathcal{V}$, with each edge as an ordered pair $(v_i,v_j)$. Here, $(v_i,v_j) \in \mathcal{E}$ indicates that $v_i$ receives information from $v_j$, so the edge direction is opposite to the information flow. Unlike undirected graphs, digraph edges are inherently oriented, and the incidence matrix can be decomposed into in- and out-incidence matrices, $E_\odot$ (non-positive entries for incoming edges) and $E_\otimes$ (non-negative entries for outgoing edges), such that $E = E_\odot + E_\otimes$, $E_\otimes E_\otimes^\top = \mathcal{D}$, and $E_\otimes E_\odot^\top = -\mathcal{A}$ \cite{8444704}. The corresponding out-Laplacian is $\mathcal{L}_\otimes = E_\otimes E^\top$. In this paper, we focus on such out-Laplacians for consensus over directed or pseudo-undirected graphs and therefore omit the $\otimes$ subscript and drop matrix arguments when the underlying graph is clear from context.

Throughout this paper, we assume graphs are connected (undirected) or contain a directed spanning tree (digraphs), since consensus is impossible otherwise \cite{mesbahi2010graph}. For a digraph, the $(i,j)$\textsuperscript{th} entry of $\mathcal{A}^k$ is nonzero if a walk of length $k$ exists from $v_i$ to $v_j$, and its properties can be analyzed via the Perron-Frobenius theorem. A matrix $\mathbf{M}$ possesses the \emph{Perron-Frobenius property} if its dominant eigenvalue $\lambda_1$ is positive with a non-negative eigenvector, and the \emph{strong Perron-Frobenius property} if $\lambda_1$ is simple, positive, and associated with a non-negative eigenvector. Finally, $\mathbf{M}$ is eventually positive if there exists $z_0>0$ such that $\mathbf{M}^z$ is positive for all $z > z_0$ \cite{noutsos2006perron}.

In many settings, edges can be assigned weights to represent the strength of information exchange between connected nodes. For a weighted digraph, the Laplacian is given by $\mathcal{L} = E_\otimes \mathcal{W} E^\top$, where $\mathcal{W}$ is a diagonal matrix of edge weights \cite{8444704}, and the adjacency matrix entries are real-valued rather than binary, with node degrees $d(v_i) = \sum_j [a_{ij}]$ forming the diagonal of $\mathcal{D}(\mathcal{G})$. Extending this idea, a pseudo-undirected graph treats an undirected edge as a pair of oppositely directed edges with potentially different weights along each direction, reflecting asymmetric interaction strengths. For an undirected graph with $m$ edges, the corresponding pseudo-undirected graph has $2m$ edges, leading to a generally non-symmetric Laplacian. 
\begin{definition}
    A pseudo-undirected graph $\mathcal{G} = (\mathcal{V},\mathcal{E},\mathcal{W})$ has a mapping $\mathfrak{F}:\mathcal{E} \rightarrow \mathbb{R}^2$ such that for each edge pair $\{v_i,v_j\}$, the weights $w_{ij}$ and $w_{ji}$ correspond to the directed edges $e_{ij}$ and $e_{ji}$, with $w_{ij} w_{ji} = 0$ only if both are zero. 
\end{definition}
In such graphs, $E_\otimes, E \in \mathbb{R}^{n \times 2m}$, and while $\mathfrak{N}(\mathcal{L}) = \mathfrak{s}\{\vmathbb{1}_n\}$ is preserved, other properties of undirected Laplacians need not hold. A pseudo-undirected graph is connected if the equivalent undirected graph (obtained by setting all non-zero edge weights to unity) is connected \cite{9147677}. In a connected pseudo-undirected graph, each node has out-degree at least one, and there exists a directed spanning subgraph $\mathcal{G}_\tau$ containing at least one globally reachable node, with a complementary subgraph $\mathcal{G}_c$ such that $\mathcal{G} = \mathcal{G}_\tau \cup \mathcal{G}_c$ \cite{8444704}. The spanning subgraph $\mathcal{G}_\tau$ contains $n-1$ directed edges, while the remaining $2m - n + 1$ edges are in $\mathcal{G}_c$. Consequently, the incidence matrix can be expressed as $E(\mathcal{G}) = [E(\mathcal{G}_\tau), E(\mathcal{G}_c)] = E(\mathcal{G}_\tau)\mathbf{R}$, where $\mathbf{R}=[\mathbf{I}_{\mathrm{n}-1}\,\vdots\,-\mathbf{I}_{\mathrm{n}-1}\,\vdots\, \mathbf{T}_\tau\, \vdots\, -\mathbf{T}_\tau]\in \mathbb{R}^{(\mathrm{n}-1)\times(2\mathrm{m}-\mathrm{n}+1)}$ encodes the relations between $\mathcal{G}_\tau$ and $\mathcal{G}_c$. Specifically, if the pseudo-undirected network has exactly twice the number of edges in $E_\tau$ (i.e., corresponding to a spanning tree), then $E = E_\tau[\mathbf{I}_{n-1}; -\mathbf{I}_{n-1}]$. For brevity, we denote $E(\mathcal{G}_\tau)$ by $E_\tau$ in the sequel.

    	\begin{lemma}[\cite{8444704}]\label{lem:TF}
		For uncertainty in the $i$-th edge, consider the uncertain edge agreement dynamics for single integrators in a rooted in-branching, $\dot{x}_\tau = -E_\tau^\top E_\otimes\left(\mathcal{W}+\mathbf{e}\Delta\mathbf{e}^\top\right)\mathbf{R}^\top x_\tau.$
		For a real, bounded, additive uncertainty, $\Delta<0$, the corresponding edge agreement protocol can be represented by the transfer function $M(s) = -\mathbf{e}^\top \mathbf{R}^\top\left[s\mathbf{I}+E_\tau^\top E_\otimes \mathcal{W} \mathbf{R}^\top\right]^{-1}E_\tau^\top E_\otimes \mathbf{e}$, 
		where $-E_\tau^\top E_\otimes \mathcal{W} \mathbf{R}^\top$ is the system matrix, $-E_\tau^\top E_\otimes \mathbf{e}$ is the input matrix, $\mathbf{e}^\top \mathbf{R}^\top$ is the output matrix, $\mathbf{e}$ is the $i$-{th} standard basis in $\mathbb{R}^{2\mathrm{m}}$, serving as an edge selection vector, and $\Delta\mathbf{e}^\top \mathbf{R}^\top x_\tau$ acts as the input to the linear system with transfer function above.
	\end{lemma}
	\begin{lemma}[\cite{8444704}]\label{lem:GM}
    Consensus over a pseudo-undirected graph is preserved if the uncertainty in a single edge weight remains below the effective gain margin of the transfer function $M(s)$, that is $|\Delta| < \frac{1}{|M(\jmath \omega_{pc})|}$, where $\omega_{pc}$ is the phase-crossover frequency corresponding to the smallest gain margin of $M(s)$.
	\end{lemma}
    \begin{problem*}
       Given the non-symmetric Laplacian induced by a pseudo-undirected graph, systematically compute the consensus value (particularly in the presence of heterogeneous and possibly negative edge weights). Additionally, investigate the admissible ranges of weight perturbations, including negative values, that preserve consensus.
    \end{problem*}

\section{Consensus over Pseudo-Undirected Graphs}
\begin{figure}[h!]
		\centering
		\includegraphics[width=\linewidth]{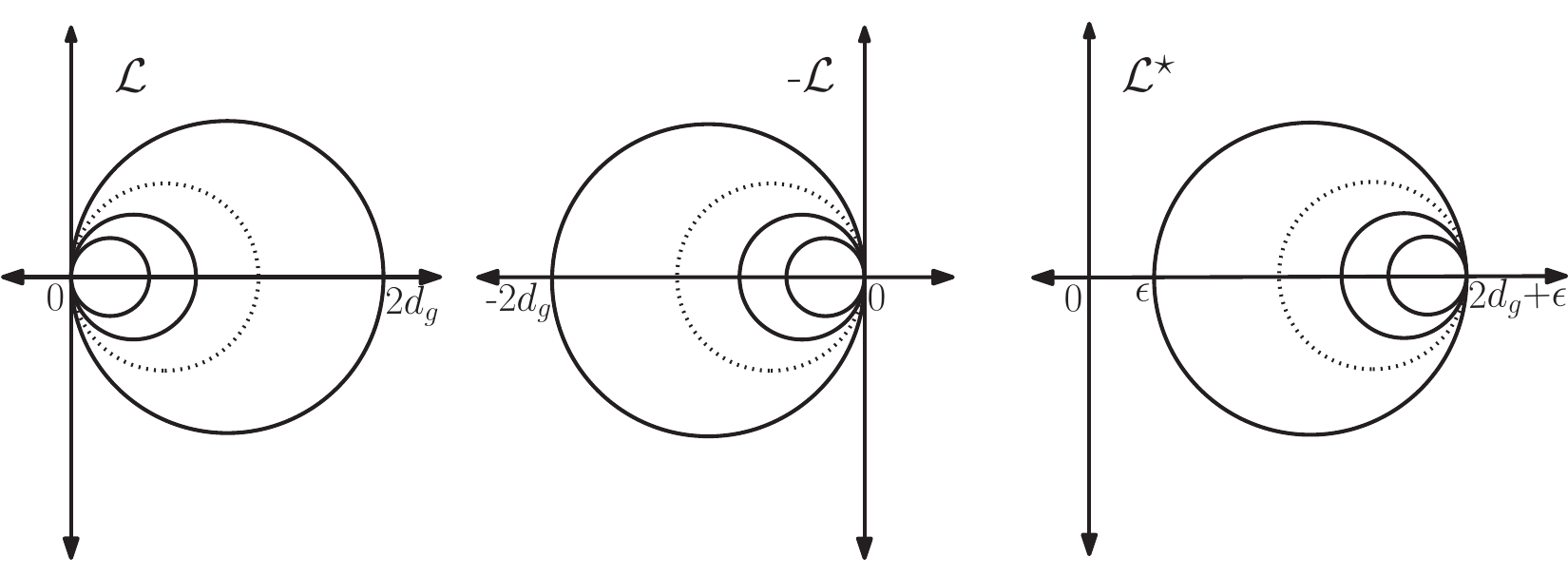}
		\caption{Ger\v{s}gorin regions for the eigenvalues of $\mathcal{L}$, $-\mathcal{L}$, $\mathcal{L}^\star$.}
		\label{fig:gdisk}
	\end{figure}

Consider the matrix $\mathcal{L}^\star = \mathcal{A} - \mathcal{D} + (2d_g + \epsilon)\mathbf{I}_n$, where $d_g = \max_i d(v_i)$ and $\epsilon > 0$. Typical Ger\v{s}gorin disks for $\mathcal{L}$, $-\mathcal{L}$, and $\mathcal{L}^\star$ are illustrated in \Cref{fig:gdisk}, assuming positive but not necessarily identical edge weights.
\begin{lemma}\label{lem:Lstar_combined}
Let $\mathcal{D}^\star = -\mathcal{D} + (2d_g + \epsilon)\mathbf{I}_n$, where $d_g = \max_i d(v_i)$ and $\epsilon>0$, and define $\mathcal{L}^\star = \mathcal{A} + \mathcal{D}^\star$. Then $\mathcal{D}^\star$ has all positive diagonal entries, and $\mathcal{L}^\star$ is eventually positive.
\end{lemma}
\begin{proof}
First, note that $\mathcal{D}^\star$ is diagonal with entries $d^\star_{ii} = 2d_g + \epsilon - d_{ii} \ge 2d_g + \epsilon - d_g = d_g + \epsilon > 0$. Hence, all diagonal entries of $\mathcal{D}^\star$ are positive. Next, consider powers of $\mathcal{L}^\star$ as 
$(\mathcal{L}^\star)^k = (\mathcal{A} + \mathcal{D}^\star)^k = \sum_{h=0}^{k} {k \choose h} \mathcal{A}^{\,k-h} (\mathcal{D}^\star)^h$. The $(i,j)$\textsuperscript{th} entry of $\mathcal{A}^{\,k-h}$ is positive if there exists a walk of length $k-h$ from $v_i$ to $v_j$, and all diagonal entries of $\mathcal{D}^\star$ are positive. Since the graph $\mathcal{G}$ has finite diameter $d$, choosing $k = d$ ensures that $(\mathcal{L}^\star)^d$ has all entries positive. Therefore, $\mathcal{L}^\star$ is eventually positive.
\end{proof}
\begin{lemma}\label{lem:ab}
If two matrices $\mathbf{A}$ and $\mathbf{B}$ are related by $\mathbf{B} = \mathbf{A} + \Delta \mathbf{I}_n$ for some $\Delta \in \mathbb{R}$, then they share the same eigenvectors, with eigenvalues shifted by $\Delta$.
\end{lemma}
\begin{proof}
Let $\mathfrak{v}$ be an eigenvector of $\mathbf{A}$ with eigenvalue $\lambda_\mathbf{A}$. Then $\mathbf{B} \mathfrak{v} = (\mathbf{A} + \Delta \mathbf{I}_n) \mathfrak{v} = \mathbf{A} \mathfrak{v} + \Delta \mathfrak{v} = (\lambda_\mathbf{A} + \Delta)\mathfrak{v}$,
so $\mathfrak{v}$ is also an eigenvector of $\mathbf{B}$ corresponding to eigenvalue $\lambda_\mathbf{A} + \Delta$.
\end{proof}
\begin{lemma}\label{lem7}
For a connected pseudo-undirected graph $\mathcal{G}$, the eigenvector of $\mathcal{L}^\top$ corresponding to $\lambda(\mathcal{L}^\top)=0$ is positive.
\end{lemma}
\begin{proof}
From \cite{noutsos2006perron}, the eventual positivity and strong Perron-Frobenius properties hold equivalently for $\mathcal{L}^\star$ and $\mathcal{L}^{\star \top}$. Hence, the eigenvectors of $\mathcal{L}^{\star \top}$ and $\mathcal{L}^\star$ corresponding to the leading eigenvalue $2d_g + \epsilon$ are positive.  Using \Cref{lem:ab}, we can write 
$\mathcal{L}^{\star \top} \mathfrak{v} = -\mathcal{L}^\top \mathfrak{v} + (2d_g + \epsilon)\mathfrak{v}$ for any eigenvector $\mathfrak{v}$ of $-\mathcal{L}^\top$ or $\mathcal{L}^{\star \top}$. For $\mathfrak{v} \in \mathfrak{N}\{\mathcal{L}^\top\}$, this gives $\mathcal{L}^{\star \top} \mathfrak{v} = (2d_g + \epsilon)\mathfrak{v}$, implying that the nullspace of $\mathcal{L}^\top$ is spanned by a positive vector, and $\mathfrak{N}\{\mathcal{L}\}\in \mathfrak{s}\{\vmathbb{1}_\mathrm{n}\}$, which is also positive.
\end{proof}
In undirected graphs, the Laplacian is symmetric, $\mathcal{L} = \mathcal{L}^\top$, and $\mathfrak{N}(\mathcal{L}) = \mathfrak{N}(\mathcal{L}^\top) = \mathfrak{s}\{\vmathbb{1}_\mathrm{n}\}$, ensuring that consensus is simply given by the average of initial states. For pseudo-undirected graphs, this symmetry is lost, and the left null vector of $\mathcal{L}$ no longer coincides with $\vmathbb{1}_n$. The problem is to compute the consensus value systematically for agents over a pseudo-undirected graph by appropriately identifying the left null vector of the Laplacian.
\begin{lemma}\label{lem:ppw}
Let $\mathbf{p}\in \mathfrak{N}(\mathcal{L}^\top)$ be a non-trivial vector for a pseudo-undirected graph with weighted Laplacian $\mathcal{L} = E_\otimes \mathcal{W} E^\top$, where $\rho(E_\otimes) = n$, $\rho(\mathcal{W}) = 2m$, and $w_{ii}>0$ for all $i$. Then the following statements are equivalent:
\begin{enumerate}
    \item $\mathbf{p} \in \mathfrak{N}(\mathcal{L}^\top)$,
    \item $(\mathbf{p}^\top E_\otimes \mathcal{W})^\top \in \mathfrak{N}(E)$,
    \item $\mathbf{v} = \mathcal{W} E_\otimes^\top \mathbf{p} \in \mathfrak{N}(E) \cap \mathcal{R}(\mathcal{W} E_\otimes^\top)$.
\end{enumerate}
\end{lemma}
\begin{proof}
By definition, $\mathbf{p}\in \mathfrak{N}(\mathcal{L}^\top)$ implies
$\mathcal{L}^\top \mathbf{p} = E \mathcal{W} E_\otimes^\top \mathbf{p} = \vmathbb{0}_n$. Transposing gives $\mathbf{p}^\top E_\otimes \mathcal{W} E^\top = \vmathbb{0}^\top_n$, which shows that $(\mathbf{p}^\top E_\otimes \mathcal{W})^\top \in \mathfrak{N}(E)$.  Since each node has out-degree at least one, $E_\otimes$ has full row rank ($\rho(E_\otimes) = n$), and $w_{ii}>0$ implies $\rho(\mathcal{W}) = 2m$, ensuring that $\mathbf{p}^\top E_\otimes \mathcal{W} \neq \vmathbb{0}^\top_{2m}$. Hence, $(\mathbf{p}^\top E_\otimes \mathcal{W})^\top \in \mathfrak{N}(E)$.  Defining $\mathbf{v} = \mathcal{W} E_\otimes^\top \mathbf{p}$, it is clear that $\mathbf{v} \in \mathcal{R}(\mathcal{W} E_\otimes^\top)$. Pre-multiplying by $E$ gives
$E \mathbf{v} = E \mathcal{W} E_\otimes^\top \mathbf{p} = \mathcal{L}^\top \mathbf{p} = \vmathbb{0}_n$, so $\mathbf{v} \in \mathfrak{N}(E) \cap \mathcal{R}(\mathcal{W} E_\otimes^\top)$. Hence, any $\mathbf{v}$ in this intersection corresponds to a $\mathbf{p}\in \mathfrak{N}(\mathcal{L}^\top)$. 
\end{proof}
By \Cref{lem:ppw}, a non-trivial vector $\mathbf{p} \in \mathfrak{N}(\mathcal{L}^\top)$ can be systematically obtained by constructing a basis for $\mathfrak{N}(E) \cap \mathcal{R}(\mathcal{W} E_\otimes^\top)$, which provides a direct method to compute the consensus value over a pseudo-undirected graph. To systematically compute $\mathbf{p} \in \mathfrak{N}(\mathcal{L}^\top)$, denote the basis sets for the nullspace and range as $\bar{U} = \{u_1, u_2, \ldots, u_{2m-n+1}\}\in \mathfrak{b}\{\mathfrak{N}(E)\}$, and $\bar{V} = \{V_1, V_2, \ldots, V_n\} \in \mathfrak{b}\{\mathcal{R}(\mathcal{W}E_\otimes^\top)\}$. By \Cref{lem:ppw}, any vector $\mathbf{v}\in\mathfrak{N}(E) \cap \mathcal{R}(\mathcal{W} E_\otimes^\top)$ can be expressed as $\mathbf{v} = \mathbf{U}x = \mathbf{V}y$, for some $x\in\mathbb{R}^{2\mathrm{m}-\mathrm{n}+1}$ and $y\in\mathbb{R}^\mathrm{n}$, where $\mathbf{U}=[u_1~\ldots ~u_{2m-n+1}]$ and $\mathbf{V}=[V_1~\ldots~ V_{n}]$.
		Since, $\rho(\mathcal{W}E_\otimes^\top) = n$ (full column rank),  a simple choice of $\mathbf{V}$ is its columns, and since $\rho(E) = n-1$, $\mathbf{U}$ can be constructed as a full-column-rank basis for $\mathfrak{N}(E)$, e.g., $$\begin{bmatrix}
			\mathbf{I}_{n-1} & \mathbf{T}_\tau & -\mathbf{T}_\tau \\
			\mathbf{I}_{n-1} & \vmathbb{0} & \vmathbb{0} \\
			\vmathbb{0} & \mathbf{I}_{m-n+1} & \vmathbb{0}\\
			\vmathbb{0} & \vmathbb{0} & \mathbf{I}_{m-n+1}\\
		\end{bmatrix}.$$ From these bases, one can compute $x = (\mathbf{U}^\top \mathbf{U})^{-1}\mathbf{U}^\top \mathbf{V}y$, and $y = (\mathbf{V}^\top \mathbf{V})^{-1}\mathbf{V}^\top \mathbf{U}x$. After substituting for $y$ in the expression of $x$, one may obtain $\mathbf{v} = \mathbf{U}x = \mathbf{P}_\mathbf{U}\mathbf{P}_\mathbf{V} \mathbf{v}$, where $\mathbf{P}_\mathbf{U}=\mathbf{U}(\mathbf{U}^\top \mathbf{U})^{-1}\mathbf{U}^\top$ and $\mathbf{P}_\mathbf{V}=\mathbf{V}(\mathbf{V}^\top \mathbf{V})^{-1}\mathbf{V}^\top$ are projection matrices. So, $\mathbf{v}$ is an eigenvector of their product and lies in $\mathfrak{N}(E) \cap \mathcal{R}(\mathcal{W}E_\otimes^\top)$. Finally, from \Cref{lem:ppw}, the corresponding $\mathbf{p} \in \mathfrak{N}(\mathcal{L}^\top)$ is obtained as $\mathbf{p}=y=(\mathbf{V}^\top \mathbf{V})^{-1}\mathbf{V}^\top \mathbf{v}$.

        Once $\mathbf{p}$ is computed, the consensus value of the multiagent systems communicating over a pseudo-undirected graph, whose dynamics is described by $\dot{\mathbf{x}}=-\mathcal{L}\mathbf{x}$, is given by ${\sum_{i}p_{i}x_i(0)}/{\sum_{i}p_{i}}$, where $p_i$ is the $i$-th entry of $\mathbf{p} \in \mathfrak{N}(\mathcal{L}^\top)$.

        In the case of strictly positive edge weights, $\mathbf{p}$ has strictly positive entries, which means that the consensus value always lies within the convex hull of the initial states $\mathcal{C}o\{\mathfrak{C}\}$ (where  $\mathfrak{C}$ is the set of the initial states). However, practical limitations exist if the consensus value must remain within $\mathcal{C}o\{\mathfrak{C}\}$. To expand the achievable range of consensus values beyond $\mathcal{C}o\{\mathfrak{C}\}$, we consider introducing negative edge weight(s), subject to bounds determined by the limiting case beyond which consensus fails.
        \begin{remark}
            Hence, given a pseudo-undirected graph with one or more negative edge weights, the next problem of interest is to determine the permissible range of these weights (or extent of perturbations an edge weight can tolerate) that guarantees convergence to consensus, and investigate how they influence the corresponding entries of $\mathbf{p}$.
        \end{remark}

    \section{Robustness Analysis for Pseudo-Undirected Path Graphs}    
        In a pseudo-undirected path, $\mathcal{P}_n$, the extreme nodes have out-degree $1$, and the rest $n-2$ vertices possess an out-degree of $2$ (refer \Cref{fig:P5}). For $\mathcal{P}_n$ with $n\geq 3$, the incidence matrix for a directed spanning sub-graph may be given as 
        \begin{equation}\label{inci}
	E_\tau =\begin{pNiceMatrix}[nullify-dots]
		1 	    &  0      &  \Cdots &    	&  0 		\\
		-1 	    &  \Ddots &  \Ddots &  		&  \Vdots	\\
		0      & \Ddots  &         &  		& 			\\
		\Vdots &  \Ddots &         &  		&  0		\\
		&         &  		&  		&  1		\\
		0      & \Cdots  &  		&  0 	& -1
	\end{pNiceMatrix}.
\end{equation}
Further, $E=E_\tau\left[\mathbf{I}_{\mathrm{n}-1}\,-\mathbf{I}_{\mathrm{n}-1}\right]$. For $\mathcal{P}_n$, the edge weight matrix is $$\mathcal{W}=\mathrm{diag}(w_{1,2},\ldots,w_{n-1,n}, w_{2,1}, \ldots, w_{n,n-1},\ldots,w_{21}),$$ and the graph Laplacian can be given by the tridiagonal matrix
\begin{equation}\label{pugeq:Lpath}
	\mathcal{L} = \begin{pmatrix}
		w_{1,2} & -w_{1,2} &  0 &  \cdots &  0 \\
		-w_{2,1} & w_{2,1}+w_{2,3} & -w_{2,3} & \ddots & 0 \\
		\vdots & \ddots &\ddots & \ddots  & \vdots\\
		0 & \cdots  &\cdots &  - w_{n,n-1} & w_{n,n-1}  
	\end{pmatrix}.
\end{equation}
\begin{figure}[ht!]
	\centering
    \resizebox{\linewidth}{!}{
	\begin{tikzpicture}
		\SetGraphUnit{3}
		\Vertex{2}
		\WE(2){1}
		\EA(2){3}
		\EA(3){4}
		\EA(4){5}
		\Edge[label = $w_{12}$](1)(2)
		\Edge[label = $w_{21}$](2)(1)
		\Edge[label = $w_{23}$](2)(3)
		\Edge[label = $w_{32}$](3)(2)
		\Edge[label = $w_{34}$](3)(4)
		\Edge[label = $w_{43}$](4)(3)
		\Edge[label = $w_{45}$](4)(5)
		\Edge[label = $w_{54}$](5)(4)
	\end{tikzpicture}%
    }
	\caption{Pseudo-undirected path, $\mathcal{P}_5$.}
	\label{fig:P5}
\end{figure}
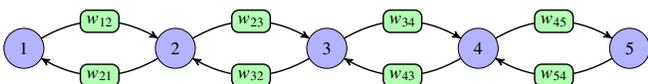
\begin{lemma}
	For $\mathcal{P}_n$ with $n\geq 3$ nodes, the entries of $\mathbf{p}\in\mathfrak{N}(\mathcal{L}^\top)$ are given by 
	\begin{equation}\label{pugeq:pipath}
		p_n = 1,~p_{n-\ell} = p_{n-\ell+1}\left(\dfrac{l_{n-\ell+1,n-\ell}}{l_{n-\ell,n-\ell+1}}\right),\,\forall\,\ell=1, 2, \ldots, n-1.
	\end{equation}
\end{lemma}
\begin{proof}
	The graph Laplacian of $\mathcal{P}_n$, for $n=5$, is given as
	\begin{equation}\label{pugeq:Lpath5}
		\mathcal{L} = \begin{pNiceMatrix}
			w_{12} & -w_{12} &  0 &  0 & 0 \\
			-w_{54} & w_{54}+w_{21} & -w_{21} &  0 & 0  \\
			0 & -w_{45} & w_{23}+w_{45} & -w_{23} &  0  \\
			0 & 0 & -w_{43} & w_{32}+w_{43} & -w_{32} \\ 
			0 & 0 & 0 & -w_{34} & w_{34}
		\end{pNiceMatrix}.
	\end{equation}
	On pre-multiplying \eqref{pugeq:Lpath5} with $\mathbf{p}$, and equating the product to $\vmathbb{0}_\mathrm{5}$, we obtain the entries of $\mathbf{p}$ for $n=5$ as
	\begin{align}
		p_1 =& \left(\dfrac{w_{34}w_{45}}{w_{12}w_{23}}\right)\left(\dfrac{w_{54}w_{43}}{w_{32}w_{21}}\right),	p_2 = \left(\dfrac{w_{34}w_{45}}{w_{23}}\right)\left(\dfrac{w_{43}}{w_{32}w_{21}}\right),\nonumber\\
		p_3 =& \left(\dfrac{w_{34}}{w_{23}}\right)\left(\dfrac{w_{43}}{w_{32}}\right),	p_4 = \dfrac{w_{34}}{w_{32}},~	p_5 =  1. \label{pugeq:ppath}
	\end{align}
	Relating the entries of $\mathbf{p}$ evaluated in \eqref{pugeq:ppath} to the Laplacian, \eqref{pugeq:Lpath5}, one can write $p_5=1,~p_4=p_5(l_{54}/l_{45}),~p_3=p_4(l_{43}/l_{34}),~p_2=p_3(l_{32}/l_{23}),~p_1=p_2(l_{21}/l_{12})$, which can be generalized to \eqref{pugeq:pipath} using mathematical induction.
\end{proof}
We now analyze the perturbation margin for $\mathcal{P}_n$ for a single edge weight. Assuming that $w_{12}$ is perturbed, $M(s)=-N(s)/D(s)$ corresponding to $e_{12}$ exhibits multiple phase-crossover frequencies, which are dependent on the choice of edge weights \cite{11018241}. In order to generalize the above result, we consider the following examples and study the transfer functions corresponding to perturbations in various edge weights, for both the pseudo-undirected graph and its underlying undirected version (where $w_{ij}=w_{ji}=1$). 

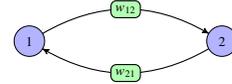
\begin{figure}[h!]
	\centering
    \resizebox{.35\linewidth}{!}{
	\begin{tikzpicture}	
		\SetGraphUnit{5}
		\Vertex{2}
		\WE(2){1}
		\Edge[label = $w_{12}$](1)(2)
		\Edge[label = $w_{21}$](2)(1)
	\end{tikzpicture}%
    }
	\caption{Pseudo-undirected path, $\mathcal{P}_2$.}
	\label{fig:P2}
\end{figure}

For simplicity of notations, let us call $-E_\tau^\top E_\otimes \mathcal{W} \mathbf{R}^\top = \mathbf{A}$, $-E_\tau^\top E_\otimes \mathbf{e}=\mathbf{B}$, and $\mathbf{e}^\top \mathbf{R}^\top=\mathbf{C}$ to compute $M(s)$ using the results in \Cref{lem:TF}. For $\mathcal{P}_2$ (see \Cref{fig:P2}), one can readily obtain $\mathbf{A} = -w_{12} - w_{21},~\mathbf{B} =-1,~\mathbf{C} =1$, thus $M(s)=-\frac{1}{s+w_{12}+w_{21}}$. For the undirected version of $\mathcal{P}_2$, $\mathbf{A} = -2,~M(s)=-\frac{1}{s+2}$ since $w_{ij}=w_{ji}=1$.
Similarly, for $\mathcal{P}_3$, one has 
$
	\mathbf{A} = \begin{bmatrix}
		-w_{12}-w_{21} & w_{23} \\
		w_{21} & -w_{23}-w_{32}
	\end{bmatrix},~ \mathbf{A}|_{w_{ij}=w_{ji}=1} = \begin{bmatrix*}[r]
		-2 & 1 \\
		1 & -2
	\end{bmatrix*}.
$
For a perturbation on $w_{12}$,
$
	\mathbf{B} =[-1,0]^\top,~\mathbf{C} =[1,0],~M(s) = -\frac{s+2}{s^2+4s+3} = -\frac{s+2}{(s+1)(s+3)}.
$
For a perturbation on $w_{23}$,
$
	\mathbf{B} =[1,-1]^\top,~\mathbf{C} =[0,1],~M(s) = -\frac{1}{s+3}. 
$
For a perturbation on $w_{32}$,
$
	\mathbf{B} =[0,1]^\top,~\mathbf{C} =[0,-1],~M(s) = -\frac{s+2}{s^2+4s+3} = -\frac{s+2}{(s+1)(s+3)}.
$
For a perturbation on $w_{21}$,
$
	\mathbf{B} =[1,-1]^\top,~\mathbf{C} =[-1,0],~M(s) = -\frac{1}{s+3}.
$

In a similar way, for $\mathcal{P}_4$, we obtain
\begin{align*}
	\mathbf{A} = &\begin{bmatrix}
		-w_{12}-w_{21} & w_{23} & 0 \\
		w_{21} & -w_{23}-w_{32} & w_{34} \\
		0 & w_{32} & -w_{34}-w_{43}
	\end{bmatrix},\\ \mathbf{A}|_{w_{ij}=w_{ji}=1} =& \begin{bmatrix*}[r]
		-2 & 1 & 0 \\
		1 & -2 & 1 \\
		0 & 1 & -2 
	\end{bmatrix*}.
\end{align*}
For a perturbation on $w_{12}$,
$
	\mathbf{B} =[-1,0,0]^\top,~\mathbf{C} =[1,0,0],~M(s) = -\frac{s^2+4s+3}{s^3+6s^2+10s+4}.
$
For a perturbation on $w_{23}$,
$
	\mathbf{B} =[1,-1,0]^\top,~\mathbf{C} =[0,1,0],~ M(s)= -\frac{s^2 + 3 s + 2}{s^3 + 6 s^2 + 10 s + 4}
    = -\frac{(s+1)(s+2)}{(s+2)(s+0.5858)(s+3.4142)}. 
$
For a perturbation on $w_{34}$,
$
	\mathbf{B} =[0,1,-1]^\top,~\mathbf{C} =[0,0,1],~M(s) = -\frac{s^2 + 3 s + 1}{s^3 + 6 s^2 + 10 s + 4}.
$
For a perturbation on $w_{43}$,
$
	\mathbf{B} =[0,0,1]^\top,~\mathbf{C} =[0,0,-1],~M(s) = -\frac{s^2+4s+3}{s^3+6s^2+10s+4}.
$
For a perturbation on $w_{32}$,
$
	\mathbf{B} =[0,1,-1]^\top,~\mathbf{C} =[0,-1,0],~
	M(s) = -\frac{s^2 + 3 s + 2}{s^3 + 6 s^2 + 10 s + 4} = -\frac{(s+1)(s+2)}{(s+2)(s+0.5858)(s+3.4142)}. 
$
For a perturbation on $w_{21}$,
$
	\mathbf{B} =[1,-1,0]^\top,~\mathbf{C} =[-1,0,0],~M(s) = -\frac{s^2 + 3 s + 1}{s^3 + 6 s^2 + 10 s + 4}.
$

Moving forward with $\mathcal{P}_5$, it follows that
$$
	\mathbf{A} = \begin{bmatrix}
		-w_{12}-w_{21} & w_{23} & 0 & 0 \\
		w_{21} & -w_{23}-w_{32} & w_{34} & 0 \\
		0 & w_{32} & -w_{34}-w_{43} & w_{45}\\
		0 & 0 & w_{43} & -w_{45}-w_{54}
	\end{bmatrix},
$$
with $\mathbf{A}|_{w_{ij}=w_{ji}=1} = \begin{bmatrix*}[r]
		-2 & 1 & 0 & 0 \\
		1 & -2 & 1 & 0 \\
		0 & 1 & -2 & 1 \\
		0 & 0 & 1 & -2 
	\end{bmatrix*}$. For a perturbation on $w_{12}$, $\mathbf{B} =[-1,0,0,0]^\top,~\mathbf{C} =[1,0,0,0]$, and
$
	M(s) = -\frac{s^3 + 6 s^2 + 10 s + 4}{s^4 + 8 s^3 + 21 s^2 + 20 s + 5}.
$
For a perturbation on $w_{23}$, $\mathbf{B} =[1,-1,0, 0]^\top,~\mathbf{C} =[0,1,0,0]$, and
$
	M(s) = -\frac{s^3 + 5 s^2 + 7 s + 3}{s^4 + 8 s^3 + 21 s^2 + 20 s + 5} = -\frac{(s+1)^2(s+3)}{(s+2)(s+0.5858)(s+3.4142)}. 
$
For a perturbation on $w_{34}$,
$
	\mathbf{B} =[0,1,-1,0]^\top,~\mathbf{C} =[0,0,1,0],~M(s) = -\frac{s^3 + 5 s^2 + 7 s + 2}{s^4 + 8 s^3 + 21 s^2 + 20 s + 5}.
$
For a perturbation on $w_{45}$,
$
	\mathbf{B} =[0,0, 1,-1]^\top,~\mathbf{C} =[0,0,0,1],~M(s) = -\frac{s^3 + 5 s^2 + 6 s + 1}{s^4 + 8 s^3 + 21 s^2 + 20 s + 5}.
$
For a perturbation on $w_{54}$,
$
	\mathbf{B} =[0,0,0,1]^\top,~\mathbf{C} =[0,0,0,-1],~M(s) = -\frac{s^3 + 6 s^2 + 10 s + 4}{s^4 + 8 s^3 + 21 s^2 + 20 s + 5}.
$
For a perturbation on $w_{43}$,
$
	\mathbf{B} =[0,0,1,-1]^\top,~\mathbf{C} =[0,0,-1,0],~
	M(s) =-\frac{s^3 + 5 s^2 + 7 s + 3}{s^4 + 8 s^3 + 21 s^2 + 20 s + 5} = -\frac{(s+1)^2(s+3)}{(s+2)(s+0.5858)(s+3.4142)}. 
$
For a perturbation on $w_{32}$,
$
	\mathbf{B} =[0,1,-1,0]^\top,~\mathbf{C} =[0,-1,0,0],~M(s) = -\frac{s^3 + 5 s^2 + 7 s + 2}{s^4 + 8 s^3 + 21 s^2 + 20 s + 5}.
$
For a perturbation on $w_{21}$,
$
	\mathbf{B} =[1,-1,0,0]^\top,~\mathbf{C} =[-1,0,0,0],~M(s) = -\frac{s^3 + 5 s^2 + 6 s + 1}{s^4 + 8 s^3 + 21 s^2 + 20 s + 5}.
$

From the preceding examples, we observe the underlying pattern in matrices $\mathbf{B}$ and $\mathbf{C}$ corresponding to various edge weights. We also observe that the transfer functions corresponding to perturbations in weights $w_{12}$ and $w_{n,n-1}$ are the same. However, in order to present the transfer function for a general case of $\mathcal{P}_n$, we note the pattern in the system matrix, $\mathbf{A}$. Based on the previous observations, it follows that the system matrix for $\mathcal{P}_n$ can be written as
\begin{equation*}
	\resizebox{\linewidth}{!}{%
		$
		\mathbf{A} = \begin{bmatrix}
			-w_{12}-w_{21} & w_{23} & 0 & \cdots & \cdots & 0 \\
			w_{21} & -w_{23}-w_{32} & w_{34} & \ddots &  & \vdots \\
			0 & w_{32} & - w_{34}-w_{43} & w_{45} & \ddots & \vdots\\
			\vdots & \ddots & \ddots & \ddots & \ddots & 0\\
			\vdots &  & \ddots & w_{n-2,n-3} & - w_{n-2,n-1}-w_{n-1,n-2} & w_{n-1,n}\\
			0 & \cdots & \cdots & 0 & w_{n-1,n-2} & -w_{n-1,n}-w_{n,n-1}
		\end{bmatrix}.
		$%
	}
\end{equation*}
For each $w_{ij}=w_{ji}=1$, the above system matrix becomes
\begin{equation*}
	\mathbf{A} = \begin{bmatrix*}[r]
		-2 &1 & 0 & \cdots & \cdots & 0 \\
		1 & -2 & 1 & \ddots & \cdots & \vdots \\
		0 & 1 & - 2 & 1 & \ddots & \vdots\\
		\vdots & \ddots & \ddots & \ddots & \ddots & 0\\
		\vdots & \cdots & \ddots & 1 & -2 & 1\\
		0 & \cdots & \cdots & 0 & 1 & -2
	\end{bmatrix*}.
\end{equation*}
Now, consider the square matrices
\begin{equation*}
	\tilde{\mathbf{B}} = \begin{bmatrix*}[r]
		-1 & 1 & 0 & \cdots & 0 \\
		0 & \ddots & \ddots & \ddots & \vdots \\
		\vdots & \ddots & \ddots & \ddots & 0 \\
		\vdots &  & \ddots & \ddots & 1 \\
		0 & \cdots & \cdots & 0 & -1
	\end{bmatrix*},~\tilde{\mathbf{C}}=\begin{bmatrix}
		1 & 0 & \cdots &  0 \\
		0 & \ddots & \ddots & \vdots \\
		\vdots & \ddots & \ddots & 0 \\
		0 & \cdots & 0 & 1
	\end{bmatrix},
\end{equation*}
with dimensions $n-1$. These matrices reveal the pattern in $\mathbf{B}$ and $\mathbf{C}$ when the path is traversed in the forward sense. The $i$-{th} column of $\tilde{\mathbf{B}}$ corresponds to $\mathbf{B}$ for the $i$-{th} perturbed edge (with $i=1,2,\ldots,n-1$), and the $i$-{th} row of $\tilde{\mathbf{C}}$ corresponds to $\mathbf{C}$ for the $i$-{th} perturbed edge.

Similarly, when the path is traversed in the opposite sense, $\tilde{\mathbf{B}}$ and $\tilde{\mathbf{C}}$ are flipped with a negative sign. Let us denote them as $\bar{\mathbf{B}}$ and $\bar{\mathbf{C}}$, such  that
\begin{equation*}
	\bar{\mathbf{B}} = \begin{bmatrix*}[r]
		0 & \cdots & \cdots & 0 & 1 \\
		\vdots &  & \iddots & \iddots & -1 \\
		\vdots & \iddots & \iddots & \iddots & 0 \\
		0 & \iddots & \iddots & \iddots & \vdots \\
		1 & -1 & 0 & \cdots & 0 
	\end{bmatrix*},~\bar{\mathbf{C}}=-\mathbf{J}_{n-1},
\end{equation*}
where $\mathbf{J}$ is the exchange matrix. Thus, we have $s\mathbf{I}-\mathbf{A}$ as
\begin{equation*}
	\resizebox{\linewidth}{!}{%
		$
		\begin{bmatrix}
			s+w_{12}+w_{21} & -w_{23} & 0 & \cdots & \cdots & 0 \\
			-w_{21} & s+w_{23}+w_{32} & -w_{34} & \ddots &  & \vdots \\
			0 & -w_{32} & s+w_{34}+w_{43} & -w_{45} & \ddots & \vdots\\
			\vdots & \ddots & \ddots & \ddots & \ddots & 0\\
			\vdots &  & \ddots & -w_{n-2,n-3} & s+w_{n-2,n-1}+w_{n-1,n-2} & -w_{n-1,n}\\
			0 & \cdots & \cdots & 0 & -w_{n-1,n-2} & s+w_{n-1,n}+w_{n,n-1}
		\end{bmatrix},
		$%
	}
\end{equation*}
and
\begin{equation*}
	\det(s\mathbf{I}-\mathbf{A}) = (s+w_{12}+w_{21})\det(s\mathbf{I}-\mathbf{A})_{(1,1)} + w_{21}\det(s\mathbf{I}-\mathbf{A})_{(2,1)}.
\end{equation*}
\begin{figure*}[ht!]
	\centering
	\begin{subfigure}{.32\linewidth}
		\centering
		\includegraphics[width=\linewidth]{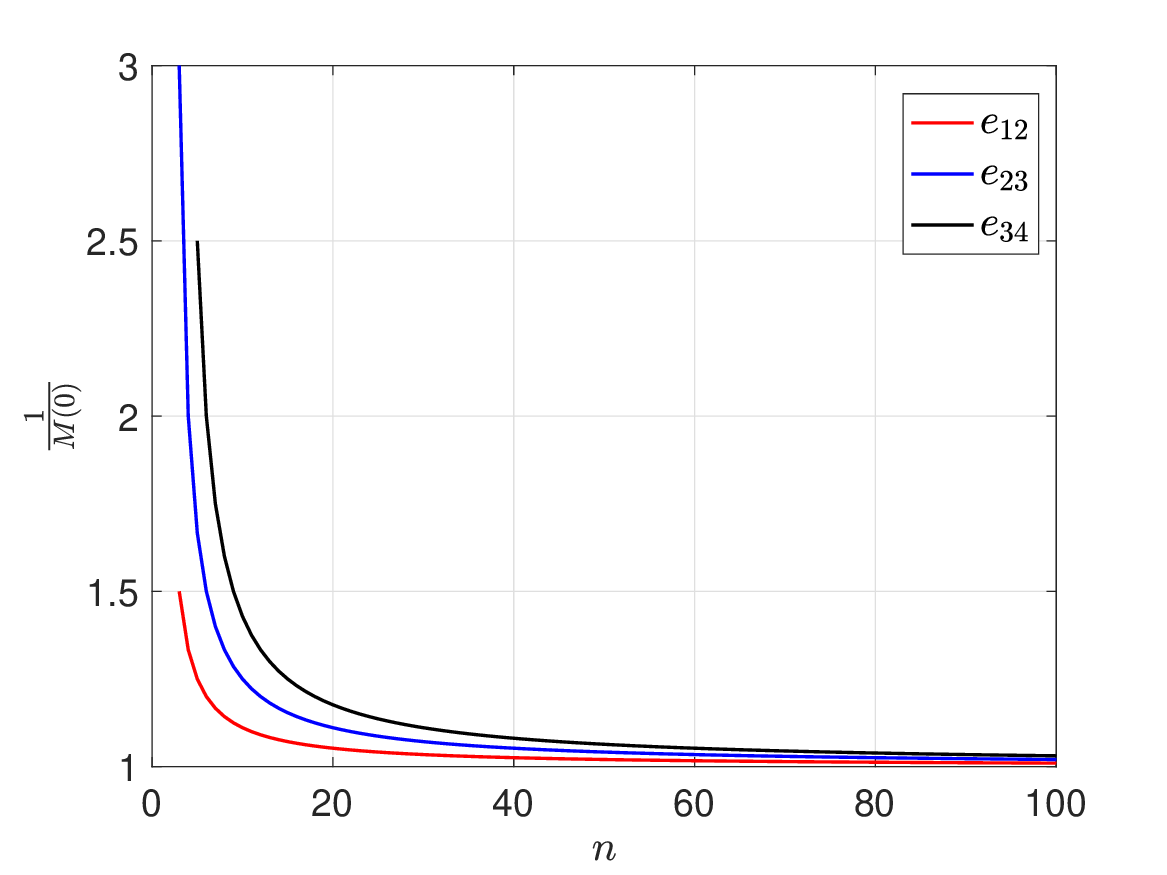}
		\caption{Variation of gain margins for perturbations in leading edges $(n\geq 3)$.}
		\label{fig:pathfwdperturbations}
	\end{subfigure}
	\begin{subfigure}{.32\linewidth}
		\centering
		\includegraphics[width=\linewidth]{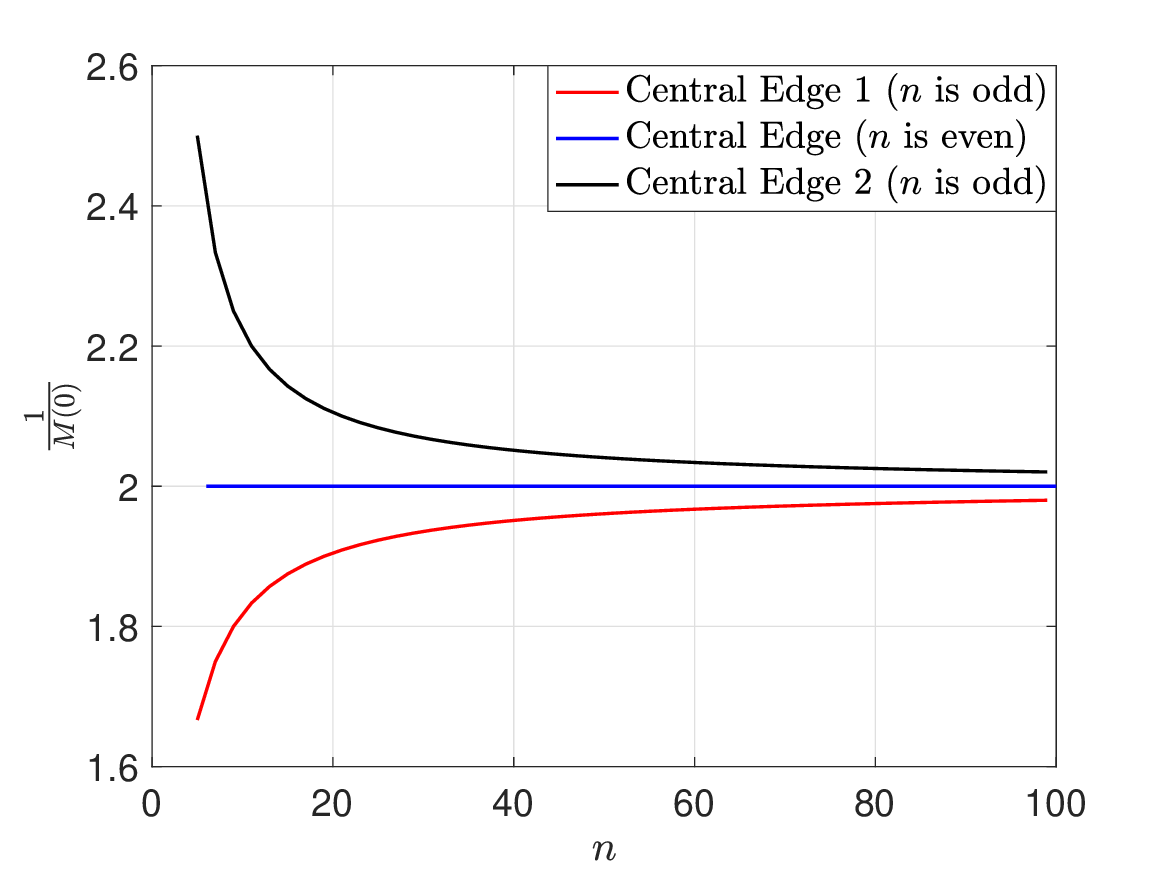}
		\caption{Variation of gain margins for perturbations in central edges $(n\geq 5)$.}
		\label{fig:pathcentralperturbations}
	\end{subfigure}
	\begin{subfigure}{.32\linewidth}
		\centering
		\includegraphics[width=\linewidth]{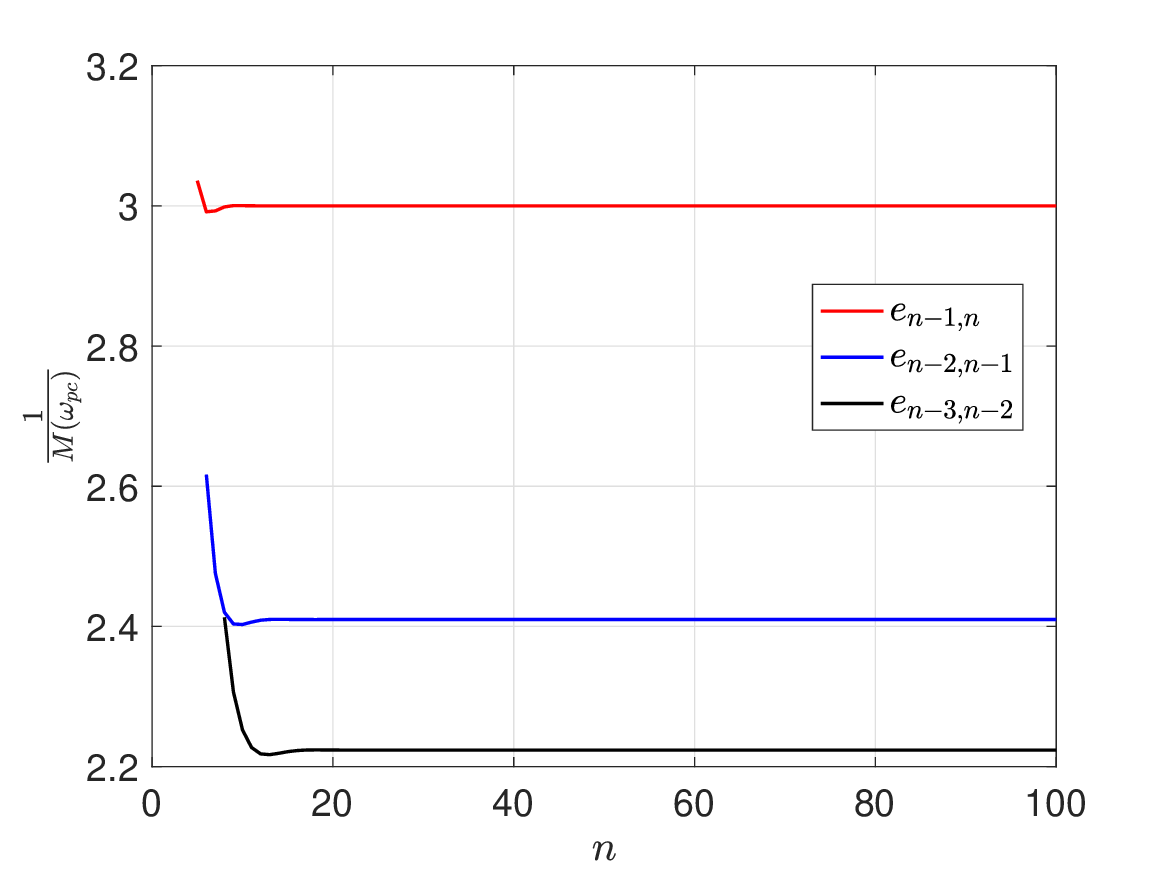}
		\caption{Variation of gain margins for perturbations in trailing edges.}
		\label{fig:pathendperturbations}
	\end{subfigure}
\caption{Variation of gain margins for perturbations in edge weights of pseudo-undirected path graphs.}
\label{fig:perturbpath}
\end{figure*}
\begin{remark}
    In a pseudo-undirected path, there are two paths in the forward and opposite sense. Due to the symmetry in its structure and in the associated matrices, $\mathbf{B}$ and $\mathbf{C}$, it suffices to evaluate the transfer functions for perturbed edges when the path is traversed in either sense. 
\end{remark}
For brevity, let us consider that the path is traversed in a forward sense.  Since the transfer functions corresponding to perturbations in $w_{12}$ and $w_{n,n-1}$ are same, it follows that
\begin{equation*}
	M(s)\vert_{w_{12},w_{n,n-1}} = -\dfrac{\det(s\mathbf{I}-\mathbf{A})_{(1,1)}}{\det(s\mathbf{I}-\mathbf{A})},
\end{equation*}
while the transfer function for any other edge in the forward path is given by
\begin{equation*}
	M(s)\vert_{\ell} = -\dfrac{(-1)^{2\ell+1}\det(s\mathbf{I}-\mathbf{A})_{(\ell-1,\ell)}-\det(s\mathbf{I}-\mathbf{A})_{(\ell,\ell)}}{\det(s\mathbf{I}-\mathbf{A})},
\end{equation*}
for $\ell=2,3,\ldots,n-1$. In general, $M(s)\vert_{w_{\ell,\ell+1}}=M(s)\vert_{w_{n-\ell+1,n-\ell}}$,  for $\ell=1,2,\ldots,n-1$. 

Due to the complex structures of $M(s)$ corresponding to different edges, and the possibility of multiple phase-crossover frequencies, an analytic computation of the gain margin may not be straightforward. However, an empirical study of the allowable bounds on the perturbation (effective gain margin) of a single edge weight brings some interesting properties to the fore. 

Note that $-M(s)$ corresponding to $\mathcal{P}_5$ has a general structure of a fourth-order transfer function, that is,
\begin{align*}
	-M(s)= \dfrac{s^3 + a_0 s^2 +a_1 s + a_2}{s^4 + b_0s^3 + b_1 s^2 + b_2 s +b_3}
	,
\end{align*}
where $a_0,a_1,a_2,b_0,b_1,b_2,b_3$ are the coefficients of various powers of the complex variable $s$. Computation of the phase-crossover frequencies requires evaluating 
        \begin{equation*}
		\resizebox{\columnwidth}{!}{%
				$
		\tan^{-1} \dfrac{\left(a_1\omega-\omega^3\right)\left(\omega^4-b_1\omega^2+b_3\right)-\left(a_2-a_0\omega^2\right)\left(b_2\omega - b_0\omega^3\right)}{\left(a_2-a_0\omega^2\right)\left(\omega^4-b_1\omega^2+b_3\right)+\left(a_1\omega-\omega^3\right)\left(b_2\omega - b_0\omega^3\right)}=-\pi,
				$%
			}
\end{equation*}
       resulting in a cubic polynomial in $\omega_{pc}^2$. In general, one phase-crossover frequency is $\omega_{pc}=0$ rad/s and the rest are functions of edge weights, that is, $\omega_{pc}=f(w_{ij})$. In general, as the number of nodes increases, the degree of the characteristic polynomial grows accordingly. Since explicit closed-form solutions for polynomials of order greater than four are not available, the analytical determination of phase-crossover frequencies in pseudo-undirected graphs becomes intractable for sufficiently large $n$. Consequently, one must resort to graphical techniques, such as the Nyquist criterion, for analysis.

To gain more insights, let us first distinctly partition the edges of a pseudo-undirected path graph as leading edges, the central edge(s), and the trailing edges. When $n$ is even, we have a single central edge, while an odd number of nodes leads to two central edges. For example, a path with 6 nodes has the third edge $(e_{34})$ as its central edge, whereas $e_{23}$ and $e_{34}$ are the central edges in a path having 5 nodes.

\Cref{fig:perturbpath} shows the variation in gain margins corresponding to edges against the number of nodes in a pseudo-undirected path graph. For instance, the gain margin corresponding to the first edge, $e_{12}$, is 1.5 for $\mathcal{P}_3$ (see \Cref{fig:pathfwdperturbations}). This value exponentially decays to 1 as the number of nodes is increased. Perturbations in other leading edges (for $n\geq 3$) show a similar variation in their respective gain margins. This essentially means that as the number of agents increases, the extent of allowable perturbation in leading edge weights reduces. Hence, even a minuscule negative perturbation can prevent the agents from reaching an agreement. \Cref{fig:pathcentralperturbations} depicts the variation of gain margins corresponding to perturbations in the central edges ($n\geq 5$ when $n$ is odd and $n\geq 6$ when $n$ is even). It is observed that the gain margin corresponding to the central edge in a path with an even number of nodes remains fixed at 2. On the other hand, the gain margins corresponding to the central edges in a path with an odd number of nodes approach this value as the number of nodes is increased. This hints at the fact that perturbations in the central edges within the allowable bound can widen the set of achievable consensus values more than that in the leading edges when the number of nodes is sufficiently large. 

It is also worth noting that perturbations in leading and central edges lead to edge transfer functions with a single phase-crossover frequency at $\omega_{pc}=0$ rad/s. From the empirical study, it follows that the gain margin pertaining to edges until the central edge in a pseudo-undirected path can be expressed as
$
\frac{1}{M(0)}= \frac{n}{n-\ell},~\ell = 1,2,\ldots,\frac{n}{2}\,(n\,\text{is even}),\, \text{or}\,\ell = 1,2,\ldots,\frac{n}{2}+1\,(n\,\text{is odd}).
$
Edges after central edges are the trailing edges. Interestingly, when these edges are subject to negative perturbations, the corresponding transfer functions exhibit multiple phase-crossover frequencies. One of them is at $\omega_{pc}=0$ rad/s while the other varies slightly when $n$ is less but becomes fixed when $n$ is large. As $n$ increases, the effective gain margin is computed using the other $\omega_{pc}$. 

As shown in \Cref{fig:pathendperturbations}, the forward edge connecting the penultimate and the last nodes has a gain margin of around 3 initially when $4<n<8$. For $n>8$, this margin becomes 3 at $\omega_{pc}=0.5774$ rad/s. For $5<n<13$, the margin reduces rapidly from $2.617$ to $2.41$ and remains fixed thereafter, if the penultimate edge is perturbed. Similarly, a perturbation on the third last edge leads to the margin dropping from $2.413$ (at $n=8$) to $2.22$ (for $n\geq15$). This behavior evidences that as we begin to perturb edges that are farther in a pseudo-undirected path graph, we have relatively higher gain margins. Unfortunately, obtaining a closed-form expression for the gain margin in this case may not be analytically tractable for large enough $n$.
\begin{remark}
	In each of the cases discussed above, the sum of the roots of the transfer functions is equal to the negative of the sum of the edge weights of the network. These transfer functions physically represent the way information flows in the network, a characterization similar to the well-known \emph{Mason's gain formula} for a linear signal flow graph.
\end{remark}

\section{Application in Cooperative Simultaneous Target Interception}
To illustrate the advantages of pseudo-undirected graphs, we consider a scenario where achieving a consensus value outside the convex hull of initial states is critical. Cooperative guidance strategies for simultaneous interception have gained attention due to benefits such as interceptor spatial separation, mission flexibility, reduced cost, and operational reliability \cite{9000526}. Achieving these benefits requires regulating interceptors’ impact times over a wide range. Existing frameworks using undirected graphs constrain the consensus to the average initial time-to-go, offering little flexibility. Directed graphs, driven only by root nodes, similarly limit achievable consensus values. The proposed pseudo-undirected framework overcomes these limitations by employing heterogeneous (possibly negative) edge weights, enabling simultaneous interception at a prescribed impact time while respecting actuator constraints.
\begin{figure}[h!]
	\centering
	\includegraphics[width=.5\linewidth]{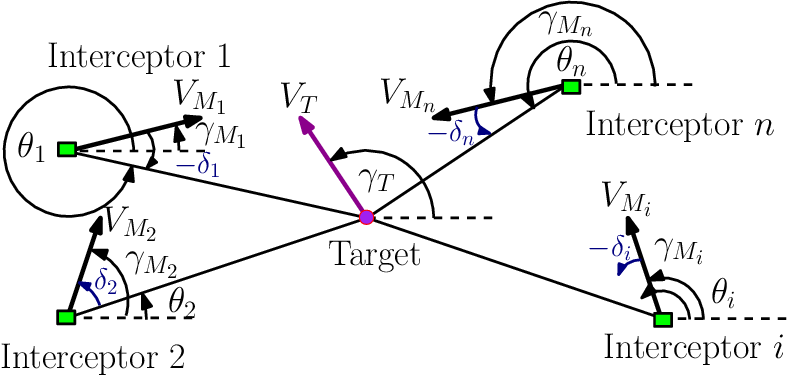}
	\caption{Planar engagement geometry for salvo interception.}
	\label{fig:eng_geo_multi}
\end{figure}

\Cref{fig:eng_geo_multi} illustrates the many-to-one planar engagement geometry where $n$ interceptors pursue a single moving target. For the $i$-{th} interceptor, $r_i$ and $\theta_i$ denote the relative range and line-of-sight (LOS) angle, while $V_{\mathrm{M}i}$ and $V_\mathrm{T}$ are the interceptor and target speeds. The interceptor’s deviation angle is $\delta_i$, its flight path angle is $\gamma_{\mathrm{M}i}$, and the target’s is $\gamma_\mathrm{T}$. The interceptor’s guidance command is the lateral acceleration $a_{\mathrm{M}i}$. To intercept a moving target, interceptors may employ a time-constrained deviated pursuit \cite{11018241} for which the cooperative guidance command can be given by $a_{\mathrm{M}_i} =V_{\mathrm{M}_i}\dot{\theta}_i - \dfrac{V_{\mathrm{M}_i}\left(V_{\mathrm{M}_i}^2 - V_\mathrm{T}^2\right) \cos^2 \delta_i}{r_i^2 \dot{\theta}_i}\sum_{j}[l_{ij}]t_{\mathrm{go}_j}$,
	where $[l_{ij}]$ are the entries of the Laplacian matrix of $\mathcal{P}_n$, and $t_{\mathrm{go}_i}$ is the time-to-go of the $i$-th interceptor guided via deviated pursuit. Each interceptor flies at $500$ m/s, while the target moves at $400$ m/s, with an initial separation of $10$ km. Due to actuator limits, the interceptors’ lateral acceleration is capped at $40$ g, where $g$ is gravity. In the trajectory plots, square markers denote interceptor initial positions, and pink circles mark the target every $10$ s. Each interceptor is labeled as $I_i$.
\begin{table}[h!]
	\caption{Edge weights for pseudo-undirected path graph.}\label{tb:wijPath}
	\centering
	\begin{tabular}{ccccccccc}
		\toprule
		$w_{12}$  & $w_{23}$ &$w_{34}$ & $w_{45}$ & $w_{21}$  & $w_{32}$ &$w_{43}$ & $w_{54}$ \\ \midrule 
		$2$ & $1.3$ & $2$ & $1.04$ & $1$ & $3.2$ & $0.15$ & $0.1$ \\  \bottomrule
	\end{tabular}	
\end{table}
\begin{table}[ht!]
	\caption{Parameters for simultaneous interception case.}\label{tb:initPath}
	\centering
		\begin{tabular}{cccccc}
		\toprule
		\phantom{-}& $I_1$  &$I_2$ & $I_3$ & $I_4$ & $I_5$\\ \midrule 
		$\theta_i$ & $0^\circ$ & $-10^\circ$& $-20^\circ$& $-165^\circ$ & $200^\circ$\\  
		$\gamma_{\mathrm{M}_i}$ & $0^\circ$ & $0^\circ$ & $0^\circ$ & $180^\circ$ & $190^\circ$\\  
		$t_{\mathrm{go}_i}(0)$ & $47.83$ s & $33.84$ s & $22.88$ s & $41.77$ s & $40.97$ s \\ \bottomrule
	\end{tabular}  
\end{table}
\begin{figure*}[h!]
	\centering
	\begin{subfigure}[t]{.33\textwidth}
		\centering
		\includegraphics[width=\textwidth]{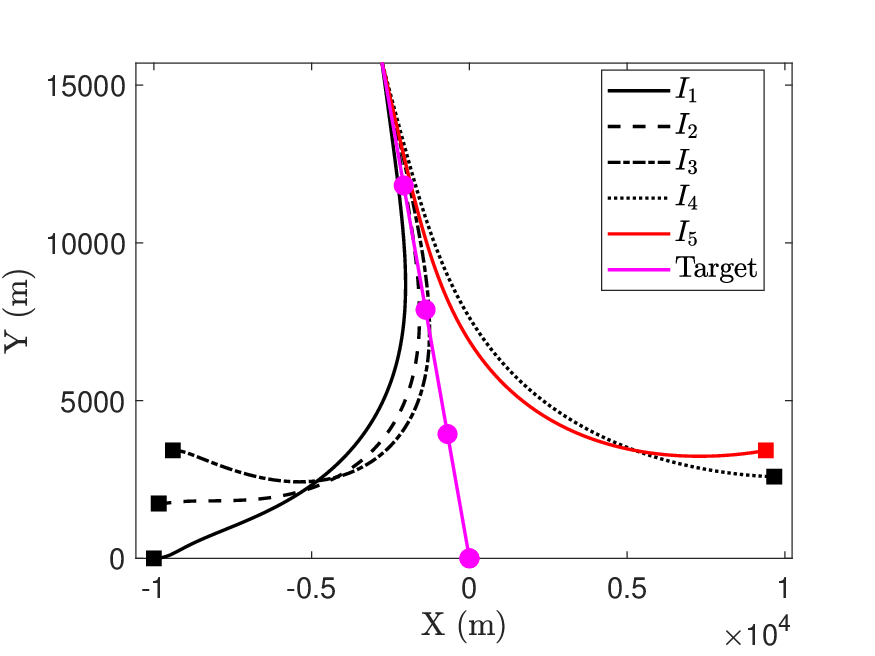}
		\caption{Trajectories.}
	\end{subfigure}
	\begin{subfigure}[t]{.32\textwidth}
		\centering
		\includegraphics[width=\textwidth]{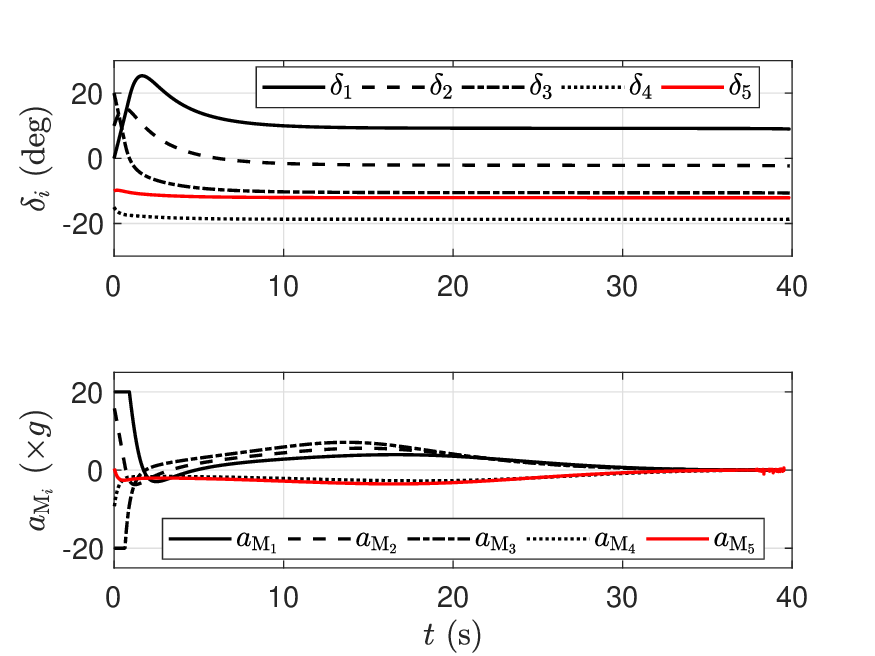}
		\caption{Deviation angles and lateral accelerations.}
	\end{subfigure}
	\begin{subfigure}[t]{.33\textwidth}
		\centering
		\includegraphics[width=\textwidth]{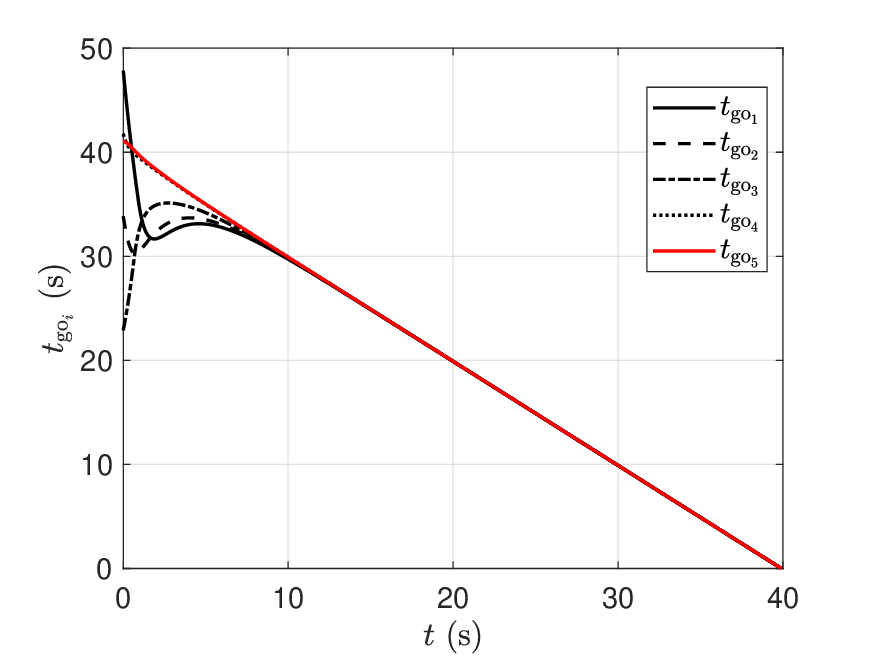}
		\caption{Time-to-go.}
	\end{subfigure}
	\caption{Weighted consensus in time-to-go.}
	\label{fig:pathHePos}
\end{figure*}
\begin{figure*}[h!]
	\centering
	\begin{subfigure}[t]{.33\textwidth}
		\centering
		\includegraphics[width=\textwidth]{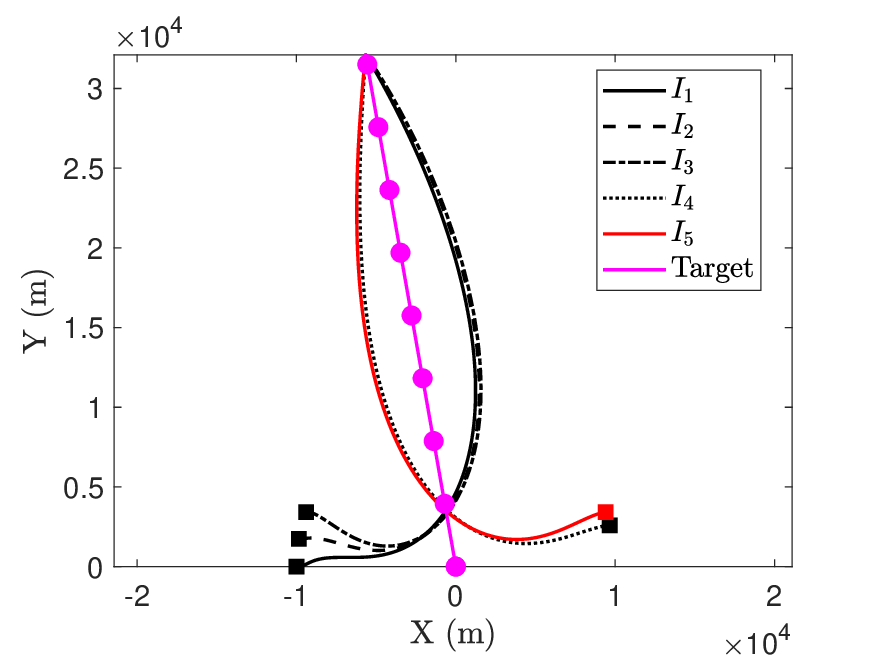}
		\caption{Trajectories.}
	\end{subfigure}
	\begin{subfigure}[t]{.32\textwidth}
		\centering
		\includegraphics[width=\textwidth]{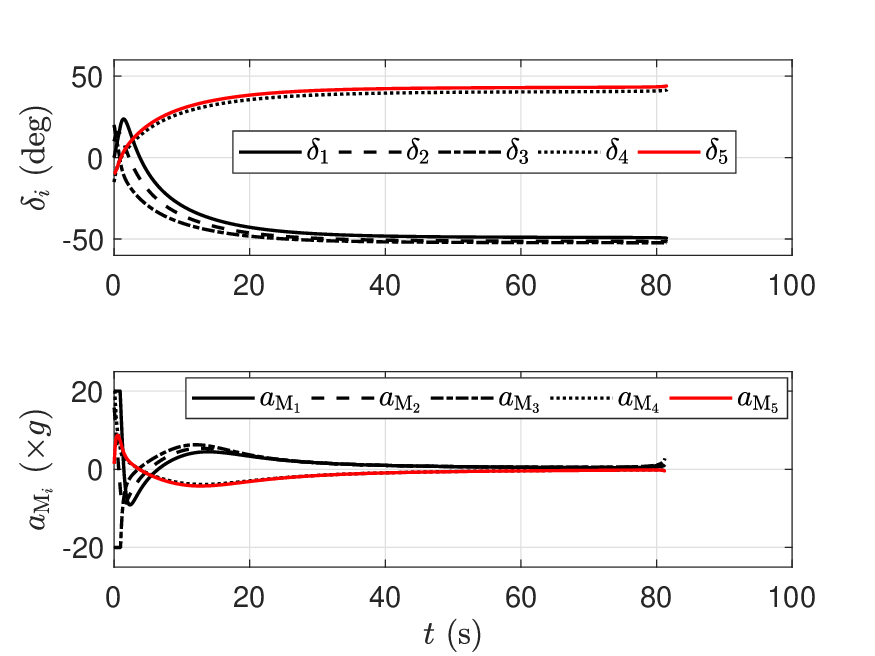}
		\caption{Deviation angles and lateral accelerations.}
	\end{subfigure}
	\begin{subfigure}[t]{.33\textwidth}
		\centering
		\includegraphics[width=\textwidth]{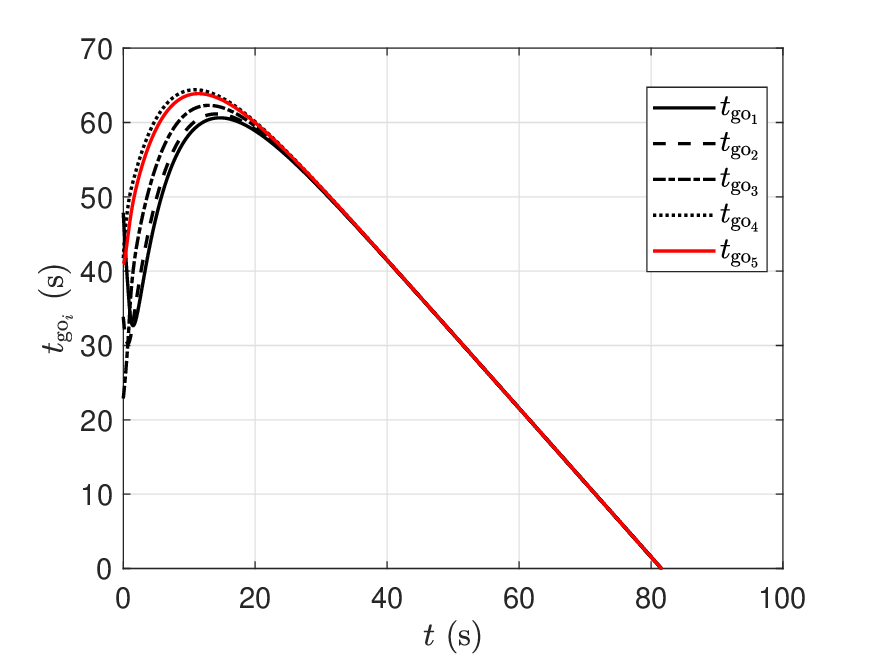}
		\caption{Time-to-go.}
	\end{subfigure}
	\caption{Impact time (consensus value) larger than the maximum initial time-to-go.}
	\label{fig:pathHeNeg}
\end{figure*}

\Cref{fig:pathHePos} shows simultaneous interception of the moving target, with $\gamma_\mathrm{T}=120^\circ$, when interceptors communicate over pseudo-undirected path graph. The edge weights for this graph are tabulated in \Cref{tb:wijPath} and the initial engagement geometry is presented in \Cref{tb:initPath}. The average value of the initial time-to-go is computed as $37.46$ s. In case of heterogeneous positive weights, the impact time is $39.86$ s (see \Cref{fig:pathHePos}). In order to intercept the target at a significantly higher impact time, we begin to perturb edge $e_{43}$. The transfer function corresponding to this edge is given by
\begin{equation}
	-M(s) = \dfrac{s^3 + 7.45 s^2 + 11.53 s + 0.63}{s^4 + 10.79 s^3 + 37.42 s^2 + 47.18 s + 14.78},
\end{equation}
having a single phase-crossover frequency at $\omega_{pc}=0$ rad/s with a gain margin of $2.0042$. Thus, by changing $w_{43}$ to $-1.1$, an impact time of $81.54$ s can be achieved. This is depicted in \Cref{fig:pathHeNeg}. Note that the achieved impact time is almost double that of the time-to-go of $I_5$, and almost $4$ times that of $I_3$. This further attests to the superiority of the proposed scheme in achieving significantly higher impact time by means of obtaining consensus outside the convex hull of the initial states. The behavior of deviation angles shows convergence to a fixed value for anticipatory maneuvers, and lateral accelerations converge to zero in the endgame.

\section{Conclusions}\label{sec:conclusion}
We presented a general framework for achieving consensus over pseudo-undirected graphs, where each edge is modeled as two oppositely directed edges with possibly different weights. We first systematically computed the consensus value by obtaining the left null vector of the Laplacian via projection-based methods. The framework accommodates heterogeneous edge weights, including negative weights within prescribed bounds, enabling flexible shaping of the consensus value beyond the average of initial node states while preserving consensus. As an application, this approach was employed for cooperative guidance of multiple interceptors, where the consensus variable corresponds to the common time-to-go. By appropriately selecting edge weights, the interceptors can achieve simultaneous interception at a desired impact time, even outside the convex hull of their initial time-to-go values, while respecting actuator limits.

				\bibliographystyle{IEEEtran}
				\bibliography{references}
\end{document}